\documentclass[prl,twocolumn,showpacs]{revtex4}

\usepackage{graphicx}
\usepackage{dcolumn}
\usepackage{amsmath}
\usepackage{amsfonts}
\usepackage{amssymb}
\begin{document}

%@twocolumnfalse\endcsname
\title{Interplay between Fermi surface topology and ordering in URu$_{2}$Si$_2$ revealed through abrupt Hall coefficient changes in strong magnetic fields
%Discontinuous change of Hall coefficient across the field-induced phase boundaries in U(Ru$_{1-x}$Rh$_x$)$_2$Si$_2$
}
\author{Y. S. Oh$^{1}$, Kee~Hoon~Kim$^{1}$, P. A. Sharma$^{2}$,
N. Harrison$^{2}$, H. Amitsuka$^{3}$, and J. A. Mydosh$^{4}$}
\address{$^{1}$CSCMR \& FPRD, School of Physics and Astronomy, Seoul National
University, Seoul, 151-747, South Korea\\
$^{2\mathrm{}}$NHMFL, Los Alamos National Lab., MS E536,
Los Alamos, NM 87545, USA\\
$^{3}$Graduate School of Science, Hokkaido University, N10W8
Sapporo 060-0810,Japan\\
$^{4}$Institute of Physics II, University of Cologne, 50937
Cologne, Germany}
%\date{\today}
%\maketitle

\begin{abstract}
Temperature- and field-dependent measurements of the Hall effect
of pure and 4~\% Rh-doped URu$_{2}$Si$_{2}$ reveal low density
(0.03 hole/U) high mobility carriers to be unique to the `hidden
order' phase and consistent with an itinerant density-wave order
parameter. The Fermi surface undergoes a series of abrupt changes
as the magnetic field is increased. When combined with existing de
Haas-van Alphen data, the Hall data expose a strong interplay
between the stability of the `hidden order,' the degree of
polarization of the Fermi liquid and the Fermi surface topology.
\end{abstract}

%]\narrowtext
\maketitle

When observed, quantum oscillatory effects such as the
de~Haas-van~Alphen (dHvA) effect provide a definitive measure of
the Fermi surface topology in metals.  There are some situations,
however, in which quantum oscillatory effects lose their
sensitivity$-$ at very low magnetic fields, through a thermally
driven phase transition or, more recently, at a quantum phase
transition. As the predictive power of Hall theory improves, the
Hall effect is becoming increasingly recognized as a viable
alternative for understanding Fermi surface changes in
$f$-electron antiferromagnets and ferromagnets tuned close to
quantum criticality~\cite{bundle,paschen}. Any knowledge of the
extent to which the $f$-electrons contribute to the Fermi surface
topology is of crucial importance for understanding the nature of
the ordering and the fate of the heavy quasiparticles.

In URu$_2$Si$_2$, Fermi surface measurements have the potential to
assist in the identification of the `hidden order' phase that
forms below $T_{\rm o}\approx$~17.5~K~\cite{palstra,chandra}. Even
if the order parameter cannot be seen directly, its effect on
modifying the Fermi surface topology could provide valuable
information on its symmetry breaking properties. Strong
correlations make this all the more challenging by both restricting the observable temperature ($T$) range
of quantum oscillations to $T\ll T_{\rm o}$~\cite{ohkuni} and
inhibiting the predictive ability of band structure calculations.
Under strong magnetic fields ($B\approx\mu_0H$),
the HO phase is destabilized through a cascade of first
order phase transitions between consecutive field-induced phases (see Fig.~1)~\cite{khkim1}. A strengthening of the correlations
in the vicinity of a putative field-tuned quantum critical
point at $B_{\rm m}\approx$~37~T~\cite{khkim2} is identified as a likely factor, although the changes in the Fermi surface topology have not been addressed.
% in increasing the impenetrability of this region of the phase diagram to quantum oscillation experiments.

In this letter, we show that Hall effect measurements extended to
high magnetic fields reveal the anomalous
contribution to become weak at low $T$. The remaining
orbital Hall effect uncovers an intricate level of interplay
between the Fermi surface topology and the stability of the
various phases of pure and 4~\% Rh-doped
URu$_2$Si$_2$~\cite{khkim2}. At low $H$ and
$T$, the enhancement of the Hall coefficient and Hall
angle~\cite{schoenes, dawson,behnia} shows that the otherwise
large Fermi surface is reconstructed into small high mobility
pockets below $T_{\rm o}$ in URu$_2$Si$_2$: a finding that is
ubiquitous among imperfectly-nested itinerant forms of broken
translational symmetry
groundstates~\cite{okajima,sasaki,chaikin,beierlein} and
consistent with low $T$ dHvA measurements on
URu$_2$Si$_2$~\cite{ohkuni}. This groundstate is then destabilized
when a magnetic field causes two of the high mobility pockets to
become spin polarized, ultimately leading to its destruction at
$\approx$~35~T. Intermediate larger and strongly polarized Fermi
surfaces appear in phases II, III and V before a fully polarized
unreconstructed Fermi surface is achieved beyond $\approx$~39~T
with 1 hole/U and 1.5~$\mu_{\rm B}$/U.

Single crystals of URu$_2$Si$_2$ and
U(Ru$_{0.96}$Rh$_{0.04}$)$_2$Si$_2$ (Rh 4 $\%$) are grown using
the Czochralski method and cut into standard Hall bar geometries
with 6-wire contacts in the tetragonal $ab$ plane. The
longitudinal and transverse voltage signals are detected
simultaneously using a pulsed magnetic field of $\sim$100~ms
duration applied along \emph{c}-axis up to 45~T. Special care is
taken to obtain Hall coefficient data under isothermal
conditions during the pulse by confirming consistency of the
resistivity and phase boundaries with measurements
made in the static magnet~\cite{khkim2}. The ability to reverse
the polarity of the pulsed magnetic field in-situ between pulses enables
the diagonal $\rho_{xx}$ and off-diagonal (Hall) $\rho_{xy}$ to be
determined accurately.
Resistivities measured at
positive and negative $B$ are subtracted and added to extract the two components.
%Temperatures
%below $\sim$~1~K are achieved using a plastic $^3$He refrigerator.
%This seems to be a main reason for the quantitative difference between $R_{\rm H}$ values presented here and those of published earlier for URu$_2$Si$_2$~\cite{bakker}.

%
\begin{figure}
\begin{center}
\includegraphics[width=0.48\textwidth]{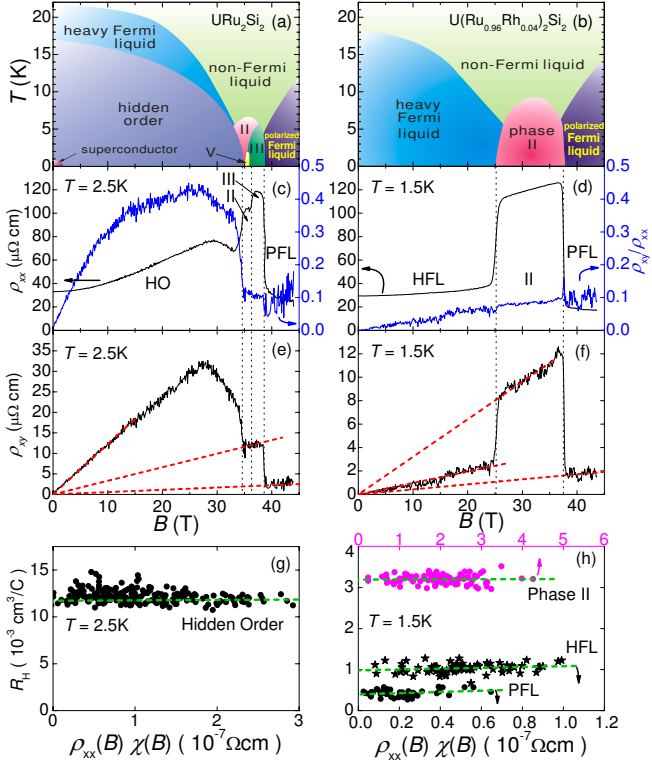}
\end{center}
\caption{Left axes: (a)\&(b) Phase diagram \cite{khkim1,khkim2},
(c)\&(d) longitudinal resistivity $\rho_{xx}$ and (e)\&(f) Hall
resistivity $\rho_{xy}$ (g)\&(h)linear extrapolations to estimate
the anomalous Hall contribution of URu$_{2}$Si$_{2}$ at $T$ = 2.5
K and U(Ru$_{0.96}$Rh$_{0.04}$)$_{2}$Si$_{2}$ at $T$ = 1.5 K,
respectively. Red dotted linear lines passing through the origin
are guides for the eye. Green dotted lines are linear
extrapolations as explained in texts. Right axes: (c)\&(d)
$\rho_{xy}$/$\rho_{xx}$ for both compounds.   } \label{Figure 1}
\end{figure}

\begin{figure}
\begin{center}
\includegraphics[width=0.47\textwidth]{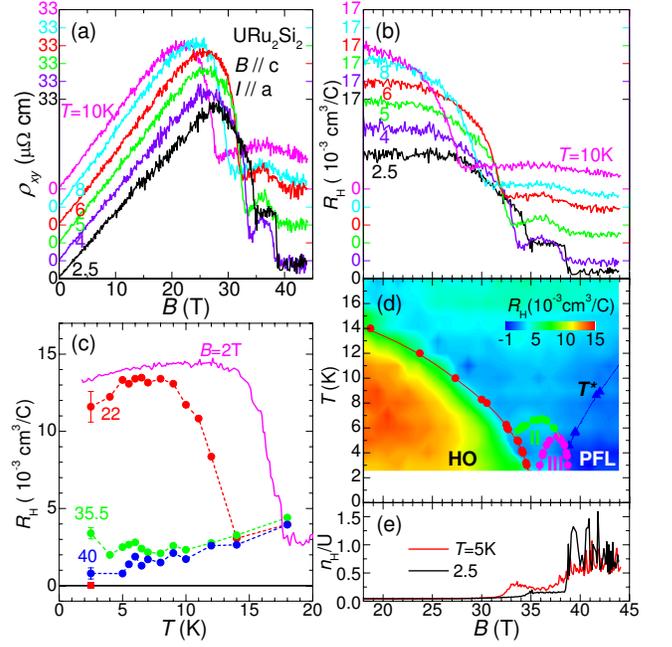}
\end{center}
\caption{ (a) $\rho_{xy}$ and (b) $R_{\rm H}~\equiv~\rho_{xy}/B$
of URu$_2$Si$_2$ at selected temperatures. (c) Temperature
dependence of $R_{\rm H}$ at selected $B$ and typical error bars
of $R_{\rm H}$ in each phase region. $R_{\rm H}$ at 2 T (magenta
line) is obtained from the temperature sweep using a
superconducting magnet. The red square represents the estimated
anomalous contribution in the HO state, which is less than 2 \% of
the measured $R_{\rm H}$. (d) The \emph{B-T} contour plot of
$R_{\rm H}$ and published phase boundaries (solid symbols)
\cite{khkim1}. (e) The effective carrier density, $n_{\rm H}\equiv
-1/eR_{\rm H}$ at $T$ = 2.5 and 5 K.} \label{fig2}
\end{figure}

\begin{figure}[tbp]
\begin{center}
\includegraphics[width=0.47\textwidth]{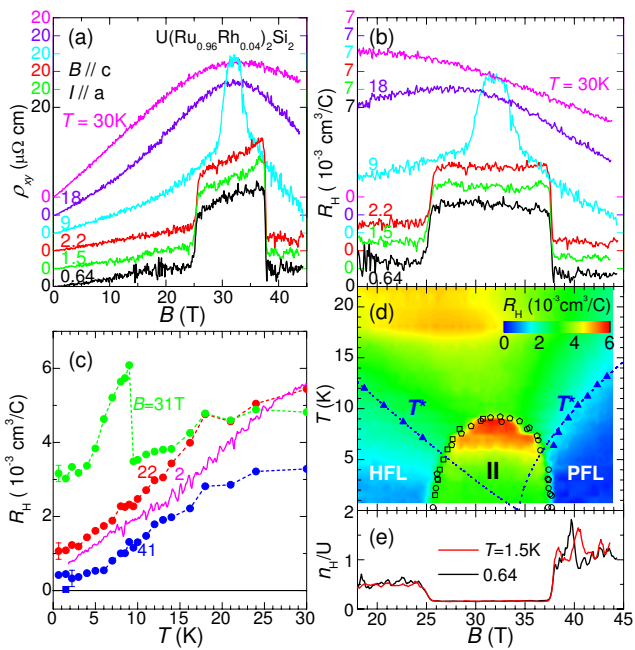}
\end{center}
\caption{(a) $\rho_{xy}$ and (b) $R_{\rm H}$ of
U(Ru$_{0.96}$Rh$_{0.04}$)$_{2}$Si$_{2}$ at selected temperatures.
(c) Temperature dependence of $R_{\rm H}$ at selected $B$ and
typical error bars of $R_{\rm H}$ in each phase region. The blue
square represents the estimated anomalous contribution in the PFL
region, which is less than 7~\% of the measured $R_{\rm H}$. (d)
The \emph{B-T} contour plot of $R_{\rm H}$ and published phase
boundaries (open circles)~\cite{khkim2}. (e) The effective carrier
density, $n_{\rm H}\equiv -1/eR_{\rm H}$ at $T=$~0.64 and 1.5~K.}
\label{fig3}
\end{figure}

Figure 1(c)-(f) shows typical $\rho_{xx}$ and $\rho_{xy}$ data for both pure and 4~\% Rh-doped URu$_2$Si$_2$ obtained at low temperatures. The $\rho_{xx}$
data are qualitatively similar to published
$\rho_{zz}$ data \cite{khkim1,khkim2}. % while the $\rho_{xy}$ data
%are consistent with that measured by Bakker {\it et
%al.}\cite{bakker} in fields of up to 40~T.
In contrast to the original Hall measurements of Bakker {\it et
al.} up to 40~T~\cite{bakker}, the present $\rho_{xy}$ data are
linear in $B$ within each high magnetic field phase, and also
within the HO phase for $B\lesssim$~26~T (we discuss the region
26~$\lesssim B\lesssim$~35~T below). This is true at all
temperatures $\lesssim$~5~K and for both samples in
Figs.~\ref{fig2}(a) and \ref{fig3}(a), enabling the Hall
coefficient $R_{\rm H}~\equiv~\rho_{xy}/B$ to be reliably
extracted in Figs.~\ref{fig2}(b) and \ref{fig3}(b). Temperature
control and isothermal conditions are important factors in
ensuring the reliability of the present measurements. The ability
to stabilize $T$ over a considerable range also enables us to
ascertain the anomalous Hall coefficient $R_{\rm aH}$ as a
relatively weak effect when compared to the orbital contribution
in the low $T$ limit.

Empirically, $R_{\rm aH}(T)\approx C_{\rm a}\rho_{xx}(T)\chi(T)$
in heavy fermion metals~\cite{paschen,fert}, where $C_{\rm a}$ is
a constant. The intrinsically narrow quasiparticle bandwidth
approaching $H_{\rm m}$ implies that even the orbital contribution
will be $T$-dependent (i.e. see Figs.~2 \& 3), thus requiring a
modified analysis. Here, we evaluate the anomalous contribution by
plotting $R_{\rm H}(B)$ versus $\rho_{xx}(B)\chi(B)$ at fixed
values of $T$ when $\chi(B)=\partial M(B)/\partial H(B)$ and
$B\approx\mu_0H$, yielding a constant slope $C_{\rm a}$ within
each phase as shown in Figs. 1(g) and 1(h).  Within the HO phase
at $T$=2.5 K, $C_{\rm a}\approx$~0.0420 T$^{-1}$ and the resultant
$R_{\rm aH}$ is less than 2~\% of the orbital contribution.
$C_{\rm a}$ cannot be extracted reliably for the other phases of
pure URu$_2$Si$_2$. However, for the 4~\% sample at $T$=1.5 K, we
obtain $C_{\rm a}\approx$~0.1013, 0.0027, and 0.1473 T$^{-1}$ for
the heavy Fermi liquid (HFL), phase II and polarized Fermi liquid
(PFL) regimes respectively, implying that $R_{\rm aH}$
contribution is always less than 7~\% of $R_{\rm H}$ for
$T\lesssim$~3~K.  This serves as an upper limit for the error
involved in extracting values of $n_{\rm H}$ from $R_{\rm H}$ at
the lowest temperatures.

In this respect, the 4~\% Rh-doped sample serves as a `reference'
sample for which there is no HO at low fields$-$ the only phase
that survives Rh-doping is phase II. The high sensitivity of the
Fermi surface transitions to perturbations in the electronic
structure may explain the complete suppression of the HO phase in
the Rh~4~\%~sample at low fields as shown in Fig.~1(b). 4~\%
Rh-doped URu$_2$Si$_2$ nevertheless has the advantage of having
its unreconstructed Fermi surface probed by Hall effect
measurements at the lowest temperature ($T\ll T_{\rm HO}$) where
the anomalous contribution is smallest. The low magnetic field
value of $R_{\rm H}$ corresponds to $n_{\rm H}\approx$~0.5 holes
per U, which is significantly larger than the additional
$\approx$~0.08 $4d$-electrons per U introduced by Rh substitution
on the Ru sites. This greatly reinforces previous assumptions that
low $T$ value of $n_{\rm H} \equiv -1/R_{\rm H}e\approx$~0.03/U
within the HO phase of pure URu$_2$Si$_2$ is primarily due to the
opening of a gap over the majority of the Fermi
surface~\cite{schoenes,dawson,behnia}. This result is robust
against differences in residual resistivity and sample quality.

One remarkable feature of the data presented in Fig.~1 is that the
tangent of the Hall angle estimated from
$\tan\theta=\rho_{xy}/\rho_{xx}$ is strongly enhanced {\it only}
within the HO phase, indicating the existence of high mobility
carriers which cause the product $\omega_{\rm c}\tau$ (where
$\omega_{\rm c}$ is the cyclotron frequency and $\tau$ is the
relaxation time) to become much larger than elsewhere in the phase
diagram. The development of high mobility carriers is a ubiquitous
feature of itinerant order parameters that break translational
symmetry (i.e. spin- and charge-and density waves) in which
imperfect-nesting leads to the creation of small pockets
~\cite{okajima,sasaki,chaikin,beierlein}. On considering the mean
free path $l=\tau v_{\rm F}$, where $v_{\rm F}=\hbar k_{\rm
F}/m^\ast$ is the orbitally averaged Fermi velocity, $k_{\rm
F}=\sqrt{2eF/\hbar}$ is the mean Fermi radius and $F$ is the dHvA
frequency, the carrier mobility becomes $\omega_{\rm
c}\tau/B\equiv e\tau/m^\ast=el/\hbar k_{\rm F}$. Thus, high
mobilities within the HO phase can be explained simply by the
creation of small pockets with small $k_{\rm F}$ values, without
needing to consider changes in $\tau$ or $l$ at $T_{\rm o}$, or
\emph{d}-wave order parameters~\cite{behnia2}.

\begin{figure}[tbp]
\begin{center}
\includegraphics[width=0.47\textwidth]{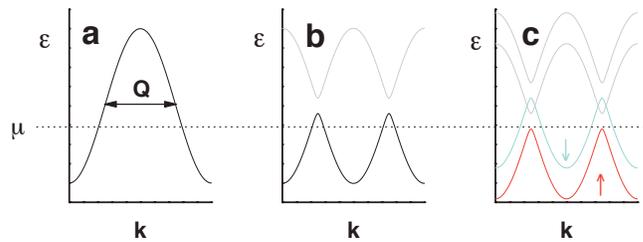}
\end{center}
\caption{Schematic cross-section in {\it k}-space though a hypothetical hole band. Broken translational symmetry ({\bf
Q}) can cause such a band (a) to be reconstructed into small pockets
with a Dirac-like dispersion (b) which are then easily polarized
by a magnetic field (c).} \label{fig4}
\end{figure}

dHvA experiments within the HO phase at $T\ll T_{\rm o}$ provide
direct evidence for small high mobility pockets~\cite{ohkuni}. The
largest of these ($\alpha$) has double the mobility
$e\tau/m^\ast\approx$~0.12~T$^{-1}$ of the others ($\beta$ and
$\gamma$) and a volume that agrees with $n_{\rm H}$ obtained from
the Hall effect measurements. Such small pockets, and subsequently
the stability of order parameter responsible for their creation,
are vulnerable to perturbations. The intrinsically large effective
masses in itinerant $f$-electron metals enables such pockets to be
easily polarized by a magnetic field, as shown schematically in
Fig.~\ref{fig4}(c). The polarization field $B_{\rm p}$ can be
estimated directly from the dHvA data by equating the Fermi
liquid, $\varepsilon_{\rm F}=\hbar eF/m^\ast$ and Zeeman
$h=g_z\sigma\mu_{\rm B}B$ (band splitting) energy scales. Here,
$F$ is proportional to the cross-sectional area of an orbit $A_k$
in reciprocal space by way of the Onsager relation $F=\hbar
A_k/2\pi e$ while $g_z\sigma\approx$~1.3 is the product of the
{\it g}-factor and spin determined from field
orientation-dependent experiments~\cite{silhanek3}. This yields
$B_{\rm p}\approx$~120, 25 and 35~T for the known $\alpha$,
$\beta$ and $\gamma$ carrier pockets, respectively~\cite{ohkuni}.
Thus, one spin component of the heaviest $\beta$ pocket, is
expected to become depopulated well within the HO phase, while
that of the $\gamma$ pocket is depopulated at the same field
$B_{\rm c}\approx$~35~T where the transition out of the HO phase
into phase II is observed (see Fig.~1 (a)). As a consequence of
such a depopulation, the chemical potential $\mu$ must
re-equilibrate itself amongst the remaining pockets, leading to
changes in their sizes and net observable changes in $R_{\rm H}$.
Indeed, as observed in Figs. 1(e) \& 2(b), once one spin component
of the heaviest $\beta$ pocket becomes depopulated beyond
$\approx$~25~T, $\rho_{xy}$, and consequently, $R_{\rm H}$ starts
a downward trend.  Since $R_{\rm H}$ is primarily due to the
$\alpha$ pocket of holes, the downward trend suggests that it is
the up spin component of a hole-like $\beta$ pocket that becomes
depopulated contributing its carriers to $\alpha$. This downward
trend then rapidly accelerates beyond $\approx$~34~T as $B_{\rm
p}$ for the $\gamma$ pocket is approached, indicating it to be
another hole pocket that becomes depopulated. Close inspection of
Fig.~2(b) reveals that the drop in $R_{\rm H}$ overshoots that
subsequently observed within phases II and III, suggesting that
depopulation of the $\beta$ and $\gamma$ pockets factors into the
destruction of the HO phase. If stability of the HO phase requires
an optimum value of translational vector ${\bf Q}$ that gaps most
of the Fermi surface, this ${\bf Q}$ will no longer be optimal
once $\mu$ begins to change, exacerbating the destabilization of
the HO phase by the Zeeman effect.

The destruction of the HO phase is accompanied by an increase in
$n_{\rm H}$ from $\sim$0.03/U inside HO to $\sim$0.15/U inside
phase II, suggesting that a significant fraction of the
unreconstructed Fermi surface is recovered. The appearance of such
a large Fermi surface will cause the Hall effect to become
dominated by lower mobility carriers, thus explaining the
restoration of $\tan\theta$ to values comparable to that of the
complete unreconstructed Fermi surface. Surprisingly, neither the
additional 0.08 electrons per U nor the disorder introduced by
charged impurities in 4~\% Rh-doped URu$_2$Si$_2$ appears to have
a detrimental effect on the stability of phase II. The collapsing
energy scale of the Fermi liquid temperature of the
unreconstructed Fermi surface associated with the field-induced
correlations could be playing a role in making this phase more
robust to disorder~\cite{khkim1,khkim2,alejandro}. This would be
especially true were the phase to involve local moment ordering of
XY electric quadrupolar degrees of freedom~\cite{silhanek4,
ohkawa}. Similarities in the value of both $R_{\rm H}$ (Figs.~2
and 3 (b) \& (c)) and $M_z$ within phase~II~\cite{neil,silhanek4}
in pure and Rh-doped URu$_2$Si$_2$ suggest that the precise degree
of polarization of the Fermi liquid is an intrinsic property of
the order parameter.

In both pure and 4~\% Rh-doped URu$_2$Si$_2$, the destruction
of the field-induced phases is followed by a complete
polarization of the unreconstructed quasiparticles bands~\cite{alejandro}
whereupon ordering is no longer possible. The saturation of the
magnetization at $\sim$~1.5$\mu_{\rm B}$/U within the polarized
Fermi liquid (PFL) state is accompanied by an increase in $R_{\rm
H}$ to a value corresponding to 0.9~$\pm$~ 0.3/U and
1.1~$\pm$~0.2/U in pure and 4~\% Rh-doped URu$_2$Si$_2$
respectively.
The realization of a
carrier density of $\approx$~1/U is consistent with the
entropy recently observed in heat capacity measurements
of the polarized bands \cite{alejandro}.

In conclusion, a clear picture of the interplay between the Fermi
surface topology, degree of polarization of the Fermi liquid and
the stability of the HO and field-induced order parameters is now
starting to emerge in URu$_2$Si$_2$. At low magnetic fields,
$T_{\rm o}$ is accompanied by a significant gapping of the
majority of the Fermi surface, leading to a reduction in the
electronic heat capacity and Hall number. An itinerant density-wave order
parameter of the type inferred from thermal conductivity
measurements~\cite{sharma} would provide a mechanism by which
small pockets of highly mobile carriers could be produced in
Fig.~\ref{fig4}. Evidence for a low density of highly mobile
carriers appears not just in the measurements of the Hall angle,
but also in the Nernst~\cite{behnia} and dHvA
effects~\cite{ohkuni}. The intrinsically heavy masses of these
pockets (compared to regular band electrons) makes them
vulnerable to polarization by a magnetic field (i.e. like in
Fig.~\ref{fig4}(c)), which subsequently destabilizes the HO phase.
The polarization of at least two of the three pockets can both be
predicted from existing dHvA measurements~\cite{ohkuni,silhanek3}
and observed directly in the Hall effect, ultimately factoring
into the destruction of the HO phase at $\approx$~35~T. Phases II,
III and V involve some form of intermediate ordering that
accommodates a partial polarization of the $f$-electrons and
partially gapped quasiparticle bands. Finally, ordering of any type is prohibited
once the unreconstructed quasiparticle bands are completely
polarized~\cite{alejandro}, yielding a simple polarized
$\approx$~1 hole/U metal at $B\gtrsim$~39~T.

%we have to be careful here
%because the tetragonal unit cell contains 2 U atoms:  I guess you
%took this into account?

%[neil wrote this for what purpose?: but then the resulting carrier
%density is not what drives the metamagnetic transition, but rather
%the density of states.]

This work is supported by Korean government through NRL program
(M10600000238) and through KRF (KRF-2005-070-C00044). YSO is
supported by the Seoul R\&BD Program, KHK by KOSEF through CSCMR,
and JAM by the A. von Humboldt Foundation. The work at NHMFL is
performed under the auspices of NSF, DOE and Florida State.


\begin{thebibliography}{40}
%\bibitem[*]{khkim} corresponding author: khkim@phya.snu.ac.kr.

\bibitem{bundle} P. Coleman, C. P\'{e}pin, Q. Si, and R. Ramazashvili, J. Phys. Condens.
Matter {\bf 13}, R723 (2001); M.~R.~Norman, Phys. Rev. B {\bf 71}, 220405
(2005); A.~Yeh {\it et al.}, Nature {\bf 419}, 459 (2002);
S.~Sachdev, Science {\bf 288}, 475 (2000); H.~L\"{o}hneysen, J.
Magn. Magn. Mat. {\bf 200}, 532 (1999); P. Gegenwart \textit{et
al.}, Phys.  Rev. Lett. {\bf 89}, 056402 (2002); G. R. Stewart,
Rev. Mod. Phys.\textbf{73}, 797 (2001); S. A. Grigera \textit{et
al}., Science \textbf{294}, 329 (2001);\emph{ibid}, \textbf{306},
1154 (2004).
\bibitem{paschen} S. Paschen {\it et al.}, Nature {\bf
432}, 881 (2004).
%% \bibitem{CeCo5In5}Y. Nakajima {\it et al.}, J. Phys. Soc. Jpn {\bf 73},
%% 5 (2004).

%\bibitem{heavyfermion} A. C. Hewson, {\it The Kondo Problem to Heavy Fermions} (Cambridge University Press, Cambridge, 1993).

\bibitem{palstra} T. T. M. Palstra {\it et al.}, Phys. Rev. Lett. {\bf 55} 2727 (1985).
\bibitem{chandra} P. Chandra {\it et al.}, Nature {\bf 417}, 831 (2002).
\bibitem{ohkuni} H. Ohkuni {\it et al.}, Philos.
Mag. B {\bf 79}, 1045 (1999).
%\bibitem{mentink} S.~A.~M.~Mentink {\it et al.}, Phys. Rev. B {\bf 53}, R6014 (1996).
%\bibitem{amitsuka1} H.~Amitsuka and T. Sakakibara, J. Phys. Soc. Jpn. {\bf 63}, 736 (1994).
%\bibitem{amitsuka2} H.~Amitsuka {\it et al.}, Phys. Rev. Lett. {\bf 83}, 5114 (1999).
\bibitem{khkim1} K. H. Kim {\it et al.}, Phys. Rev. Lett. {\bf 91},
256401 (2003).
\bibitem{khkim2} K. H. Kim {\it et al.}, Phys. Rev. Lett. {\bf 93}, 206402 (2004).
\bibitem{schoenes} J. Schoenes {\it et al.}, Phys. Rev. B {\bf 35}, 5375
(1987).
\bibitem{dawson} A.~LeR Dawson {\it et al.}, J. Phys.: Cond. Matt. {\bf 1}, 6817 (1989).
\bibitem{behnia} R. Bel {\it et al.}, Phys. Rev. B {\bf 70}, 220501(R)(2004).
\bibitem{okajima} K.~Okajima and S.~Tanaka, J. Phys. Soc. Jpn {\bf 53}, 2332 (1984).
\bibitem{sasaki} T.~Sasaki, S.~Endo and N.~Toyota, Phys. Rev. B {\bf 48}, 1928 (1993).
\bibitem{chaikin} P.~M.~Chaikin, J. Phys. I (France) {\bf 6}, 1875 (1996).
\bibitem{beierlein} U.~Beierlein {\it et al}, Synth. Met. {\bf 103}, 2593 (1999).

\bibitem{bakker} K. Bakker {\it et al.}, Physica B {\bf
186-188}, 720 (1993).

\bibitem{fert} A. Fert and P. M. Levy, Phys. Rev. B {\bf36}, 1907
(1987).
\bibitem{behnia2} K. Behnia {\it et al.}, Phys. Rev. Lett. \textbf{94}, 156405(2005).
%\bibitem{hybrid}...
%\bibitem{comment2}The estimation of $C_{\rm B}$ in the other regions of URu$_2$Si$_2$ cannot be made reliably because of the narrow region of phase II and small value of $R_{\rm H}$ in the PFL region.
%\bibitem{tempano}Further estimation as a function of $T$ by use of the same value of $C_{\rm B}$ determined at lowest temperatures shows that the anomalous contribution increases slowly but always remains less than 10 \% below 10~K inside HO of URu$_2$Si$_2$, and phase II \& HFL regions of the Rh~4~\%~sample, while in the PFL region of the Rh~4~\%~sample, it increases to about 30 \% at 10~K.
\bibitem{silhanek3} A.~Silhanek {\it et al.}, Physica B {\bf 378-380}, 373 (2006).

\bibitem{alejandro} A.~V. Silhanek {\it et al.}, Phys. Rev. Lett. {\bf 95}, 026403 (2005).
\bibitem{silhanek4} A.~V.~Silhanek {\it et al.}, Phys. Rev. Lett. {\bf 96}, 136403 (2006).
\bibitem{ohkawa} F.~Ohkawa and H.~Shimizu, J. Phys.: Cond. Matt. {\bf 11}, L519 (1999).
\bibitem{neil} N. Harrison, M. Jaime, and J. A. Mydosh, Phys.  Rev. Lett. {\bf 90}, 096402 (2002).
\bibitem{sharma} P. A. Sharma {\it et al.}, Phys. Rev. Lett. \textbf{97}, 156401
(2006).


%\bibitem{kiss} A. Kiss and P. Fazekas, Phys. Rev. B {\bf 71}, 054415
%(2005).
% \bibitem{sharma} P. A. Sharma {\it et al.}, Phys. Rev. Lett. \textbf{97}, 156401 (2006); K. Behnia {\it et al.}, \emph{ibid.} \textbf{94}, 156405(2005).
%\bibitem{gorkov} V. Barzykin and L. P. Gor'kov, Phys. Rev. Lett. {\bf 70}, 2479 (1993).
%\bibitem{santini} P. Santini and G. Amoretti, Phys. Rev. Lett. {\bf 73}, 1027 (1994).
% \bibitem{ikeda} H. Ikeda and Y. Ohashi, Phys. Rev. Lett. {\bf 81}, 3723 (1998).
% \bibitem{chandra} P. Chandra, P. Coleman, J. A. Mydosh, and V. Tripathi, Nature {\bf 417}, 831 (2002).
% \bibitem{buyers} W. J. L. Buyers, Physica B {\bf 223-224}, 9-14 (1996).

\end{thebibliography}
\end{document}